\documentclass[12pt,preprint]{aastex}

\shorttitle{ALMA view on protoplanetary disks}
\shortauthors{Semenov et al.}

\begin{document}

\title{Chemical and thermal structure of protoplanetary disks as observed with ALMA}

\author{D. Semenov, Ya. Pavlyuchenkov, Th. Henning, S. Wolf, and R. Launhardt}
\affil{Max Planck Institute for Astronomy, K\"onigstuhl 17, 69117 Heidelberg, Germany}
\email{semenov,pavyar,henning,swolf@mpia.de}

% \author{D. Semenov}
% \affil{Max Planck Institute for Astronomy, K\"onigstuhl 17, 69117 Heidelberg, Germany}
% \email{semenov@mpia.de}
%
% \author{Ya. Pavlyuchenkov}
% \affil{Max Planck Institute for Astronomy, K\"onigstuhl 17, 69117 Heidelberg, Germany}
% \email{pavyar@mpia.de}
%
% \author{Th. Henning}
% \affil{Max Planck Institute for Astronomy, K\"onigstuhl 17, 69117 Heidelberg, Germany}
% \email{henning@mpia.de}
%
% \and
%
% \author{S. Wolf}
% \affil{Max Planck Institute for Astronomy, K\"onigstuhl 17, 69117 Heidelberg, Germany}
% \email{swolf@mpia.de}

\begin{abstract}
We predict how protoplanetary disks around low-mass young stars
would appear in molecular lines observed with the ALMA interferometer.
Our goal is to identify those molecules and transitions that can be
used to probe and distinguish between chemical and physical disk structure
and to define necessary requirements for ALMA observations.
Disk models with and without vertical temperature gradient
as well as with uniform abundances and those from a chemical network are considered.
As an example, we show the channel maps of HCO$^+$(4-3) synthesized with a non-LTE line
radiative transfer code and used as an input to the GILDAS ALMA simulator to produce
noise-added realistic images. The channel maps reveal complex asymmetric patterns even
for the model with uniform abundances and no vertical thermal gradient.
We find that a spatial resolution of $0.2-0.5\arcsec$ and 0.5--10~hours of integration time
will be needed to disentangle large-scale temperature gradients and the chemical
stratification in disks in lines of abundant molecules.
\end{abstract}

\keywords{accretion, accretion disks --- astrochemistry ---
circumstellar matter --- molecular processes --- planetary
systems: protoplanetary disks --- instrumentation: interferometers}

\section{Introduction}
The cradles of planet formation -- protoplanetary disks -- are ubiquitous around young stars, at least
of low and intermediate masses ($\la 8\,M_\sun$).
It is imperative to study disk structure and evolution in detail if one aims to understand initial
conditions for planet formation.
These objects are best studied at (sub-)millimeter wavelengths in thermal dust continuum and
various molecular lines tracing their physical and chemical structure. However, being small ($\sim 200-1\,000$~AU)
and of relatively low mass ($\sim 0.01{\rm M}_\sun$), such disks are hard to study observationally
as high spatial resolution and sensitivity are required.
To date only several nearby disks around T Tauri and Herbig Ae/Be stars have been
spatially resolved in a handful of molecular lines \citep{Bea07}. These observational data
coupled to advanced modeling allowed us to constrain basic disk parameters such as size, mass,
temperature, density distribution, kinematics, and ionization structure
\citep[e.g.,][]{Sea05,Qi_ea06,Pietu_etal2007,Dutrey_ea07}.
An interesting result concerns the vertical temperature gradient that seems to be present
in some disks (e.g., \object{DM Tau}, \object{AB Aur}) while absent in others (e.g., \object{LkCa 15}),
see \citet{Pietu_etal2005,Pietu_etal2007}.

The situation will change dramatically when the Atacama Large Millimeter Array (ALMA), equipped with
50 12-m antennas, will come into operation in 2012. ALMA will be capable of
imaging protoplanetary disks at spatial resolutions up to $\sim 0.04-0.005\arcsec$ in a frequency range
between 100 and 950~GHz. This will allow direct detection and characterization of disk instabilities,
resulting in clumpy structures (vorticities, ``spiral arms'', etc.) as well as inner gaps and holes
induced by forming giant planets \citep{Wolf_ea02a,Wolf_ea02b,WDA_05,Nea06}. The large-scale chemical structure
of protoplanetary disks will become accessible in rotational lines of many abundant molecules \citep{ARea07,Kea07}.
As demonstrated by \citet{Pea07}, excitation of molecular lines in disks with strong
gradients of physical conditions and chemical structure can be a complicated process and may be hard to interpret.
% on spectral line images.

The main aim of this work is to unravel the potential of ALMA
to distinguish between various temperature and chemical effects in protoplanetary disks, with
an emphasis on discerning spatial resolution and integration time needed for this goal.
Such observations will allow us to fully explore the validity of state-of-the-art disk
chemical models on a global scale \emph{for the first time}, and advance our understanding of the
disk physics enormously.
In contrast to previous studies of \citet{Nea06} and \citet{Kea07}, we take into account both the non-LTE
excitation and realistic chemical structure.
%_______________________________________________________________
%
\section{Disk models}
\begin{figure}
%\epsscale{0.85}
\plotone{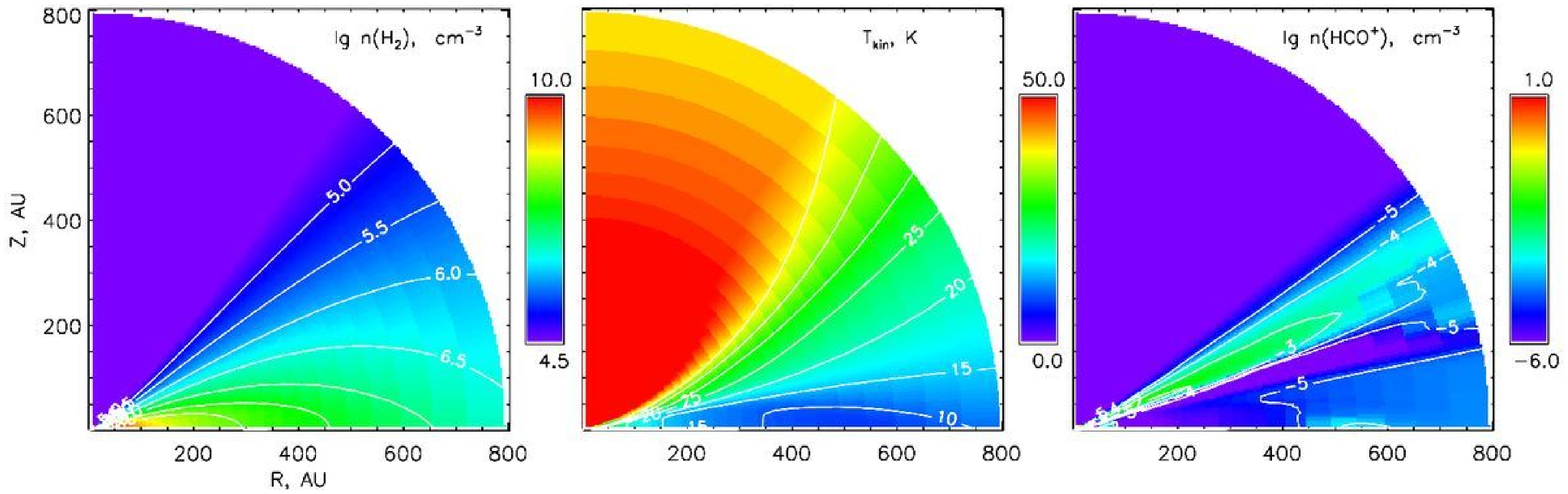}
\caption{Distributions of the particle density (left panel), temperature (middle panel), and absolute
abundances of HCO$^+$ at 5~Myr (right panel) for the model most similar to the DM~Tau disk.}
\label{disk_struc}
\end{figure}

%_______________________________________________________________
We adopt a steady-state flared disk model with
a radius of 800~AU, an accretion rate
$\dot{M}=10^{-8}\,M_\sun$\,yr$^{-1}$, a viscosity parameter
$\alpha = 0.01$, and a mass of $M=0.07\,M_\sun$ \citep{pad1999}.
The disk has a microturbulent velocity
of 0.1~km\,s$^{-1}$ and a Keplerian rotation profile,
$V(r) \propto r^{-1/2}$.
This disk model resembles the disk around the young T Tauri star DM~Tau located at the distance of
$\approx 140$ pc \citep{Dutrey_ea07}.
Its thermal and density structure is shown in Fig.~\ref{disk_struc} (left and middle panels).
To mimic the case when the vertical temperature gradient is weak or absent,
we also consider the same disk model, but with a vertical temperature distribution
fixed to the value at one pressure scale height.
% ($Z/R \approx 0.2$).

Using both disk models, 5~Myr of disk chemical evolution is modeled with a
gas-grain chemical model from \citet{Vea07}.
The gas-phase reaction data are adopted from the RATE\,06 database \citep{rate06}.
Ionization and dissociation of the gas-phase species as well as desorption of surface species due to
cosmic rays, stellar X-rays, stellar and interstellar UV radiation are taken into account.
The disk is illuminated by UV radiation from the central star with
an intensity $G=410\,G_0$ at $r=100$~AU and by interstellar UV radiation with
intensity $G_0$ \citep[][]{Draine_78,van_Dishoeck88}. The X-ray ionization rate in a given disk region
is computed according to \citet{Glassgold_ea97a}.
The adopted cosmic-ray ionization rate is 1.3$\times$10$^{-17}$~s$^{-1}$ \citep{ST68}.
Molecules are assumed to stick to grains with 100$\%$
efficiency. We also consider dissociative recombination of ions on charged dust grains and grain
re-charging, with a dust-to-gas mass ratio of $1\%$. To model surface chemistry on the $0.1\mu$m olivine
grains, a set of surface reactions from \citet{HHL92} and \citet{HH93} is used.
% and desorption energies from \citet{GH_06} are used.

In what follows, we will concentrate on HCO$^+$ -- a representative abundant molecule that traces
the ionization fraction and possesses strong rotational transitions, and which is readily observed
in disks \citep{Dutrey_ea07}.
The modeled distribution of the HCO$^+$ absolute abundances
(molecules\,cm$^{-3}$) in the DM~Tau disk at 5~Myr is shown in Fig.~\ref{disk_struc} (right panel).
A layered chemical structure with the maximal HCO$^+$ concentration reached at intermediate disk
heights is clearly visible. Many observed species have a similar stratification,
e.g., CO, CS, HCN, etc. \citep{AH99,Vea07}.

% Often for simple analysis of the observational data molecular abundances are assumed to be constant
% with respect to hydrogen.
Strong turbulent mixing and/or global advection flows can smooth these abundance gradients, leading
to more uniformly distributed abundances \citep{Willacy_ea06,Semenov_ea06}.
Therefore, we also consider a model with an uniform HCO$^+$ abundance of $10^{-9}$ relative to H$_2$.
Thus, our study is based on three disk models: % of various degree of complexity:
(1) the model with chemical stratification and vertical temperature gradient, (2) the same model but with
uniform abundances, and (3) the model with uniform abundances and no vertical temperature gradient.

%--------------------------------------------
\section{Synthetic channel maps}
\begin{figure}
\epsscale{0.85}
\plotone{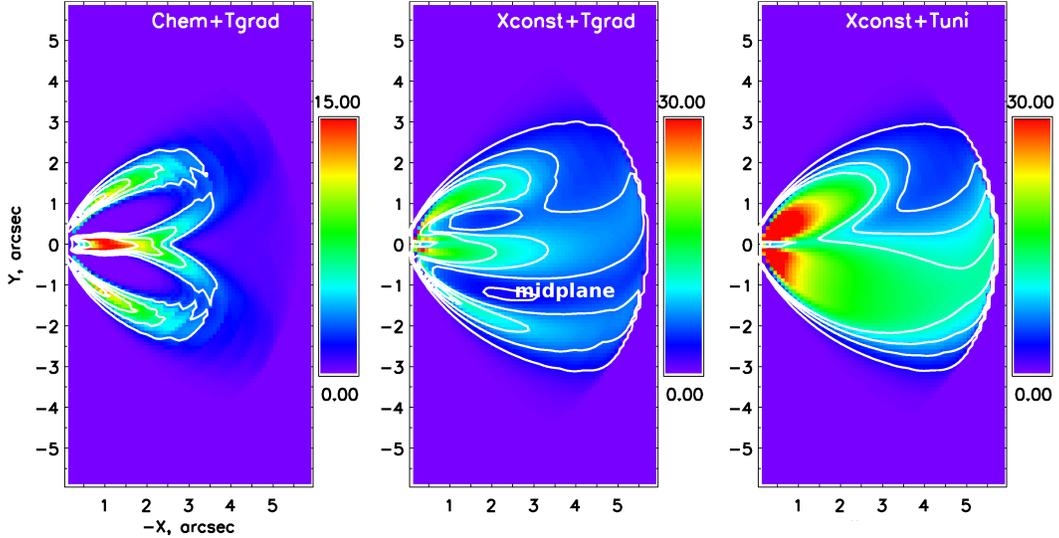}
\caption{(Left to right) The continuum-subtracted HCO$^+$(4-3) synthetic map at
the $V=+0.77$~km\,s$^{-1}$ velocity channel
for the three disk models: the model with chemical stratification and temperature gradients,
the same model but with the uniform abundances, and the model with the uniform abundances and
no vertical temperature gradient. The inclination angle is $60\degr$. Intensity is given in units of
radiative temperature (Kelvin).}
\label{chan_maps_3}
\end{figure}

%--------------------------------------------
Using these three DM~Tau-like disk models, and the 2D non-LTE line radiative transfer code of
\citet{Pea07} with thermal dust continuum included, we synthesize ``ideal'' (beam-unconvolved) channel maps.
The collisional rates are taken from the Leiden Atomic and Molecular
Database\footnote{\url{http://www.strw.leidenuniv.nl/~moldata/}} \citep{LAMDA}, while dust opacities for silicate grains
are from \citet{DL84}. At the $60\degr$ inclination the integrated HCO$^+$(4-3) line width is about 3~km\,s$^{-1}$,
with a peak intensity reached at about 0.8~km\,s$^{-1}$ \citep[see Fig.~10 in][]{Pea07}.
The corresponding $0.1$~km\,s$^{-1}$ wide velocity channel at
$V=0.77$~km\,s$^{-1}$ of the continuum-subtracted
HCO$^+$(4-3) map is shown in Fig.~\ref{chan_maps_3}.
%\footnote{(An animation showing all channel maps is available in the electronic
%edition of the Journal).}.

This close to edge-on orientation is particularly favorable as the layered and uniform disk chemical structures
can be clearly distinguished in the channel maps, if sufficiently small channel widths of $0.1-0.2$~km\,s$^{-1}$
are used. All synthetic channel maps reveal a complex
pattern and are asymmetric. The model with both non-zero temperature and abundance gradients has a remarkable
``omega'' shape, where the cold midplane with low HCO$^+$ concentration appears as two intensity ``holes''
(Fig.~\ref{chan_maps_3}, left panel). A steep temperature gradient in radial direction toward the inner disk region
is clearly visible, while the vertical temperature gradient cannot be fully traced with the HCO$^+$ lines
due to strong chemical stratification of this species.
The HCO$^+$(4-3) emission is thermalized and optically thin in this case.

The two models with uniform abundances have relatively high HCO$^+$ column densities so that
self-absorption becomes important. In the model with fixed vertical temperature this effect leads
to a layer with low intensity in the upper part of the map (Fig.~\ref{chan_maps_3}, right panel).
However, the overall
intensity of the HCO$^+$(4-3) emission is higher compared to the model with layered abundances. The radial
temperature gradient is also prominent in this case.

The model without chemical stratification but with vertical temperature gradient also reveals the self-absorption
layer in the upper part of the channel map. The excitation temperature of the HCO$^+$(4-3) transition
changes strongly with the disk height and has the lowest value in the midplane. This zone of low intensity
appears as a fake ``spiral arm'' in the lower part of the map (marked as ``midplane'' in Fig.~\ref{chan_maps_3}, middle panel).
Note that such a pattern somewhat resembles the ``omega'' structure for the model with chemical
stratification of HCO$^+$. While we show only one representative channel, one should bear in mind that
analysis of the interferometric data is more reliable when all channels and dust continuum emission
are taken into account.
% The combination of multi-wavelength continuum and line channel maps is a useful way
% to distinguish between details of disk density distribution and other parameters (e.g., temperature,
% abundances, velocity).

%--------------------------------------------
\section{ALMA observations}
%--------------------------------------------
Despite the fact that all the models considered can be distinguished
in the synthetic channel maps, it is important to study whether
ALMA is able to disentangle the effects of thermal and chemical structure in disks.
To simulate the observations with ALMA, we use the synthetic HCO$^+$(4-3) channel maps as an input
to the GILDAS simulation software of \citet{memo398} (version from October 2007).
A position close to DM~Tau is assumed: $\alpha_{\rm 2000}=4$h~33m~48s, $\delta_{\rm 2000}=18\degr$~$10\arcmin$~$09\arcsec$
\citep{Ducourant_ea05}.
%Unfortunately, this package is not yet capable of simulating
%spectral line observations in all channels in one run, so we have to focus on the $0.77$~km\,s$^{-1}$ channel
We focus on the $0.77$~km\,s$^{-1}$ channel
(with 120~kHz bandwidth or $0.1$~km\,s$^{-1}$ velocity width). Array configurations with 64 antennas
built in GILDAS are considered. Note that in reality ALMA will consist of 50 antennas with different
configurations, so our analysis will tend to underestimate required observing time by a factor of 1.5
but the simulated spectra will not be significantly affected.

Typical weather conditions at
the Chajnantor plateau are assumed \citep[see][]{memo398}, with the following main types of errors leading to noise:
1) receiver temperature is 80~K, system temperature is 230~K \citep[at $\nu = 357$~GHz, see][]{memo393},
2) random pointing errors during the observation are $0.6\arcsec$,
3) relative amplitude errors are 3\% with a 6\%~hour$^{-1}$ drift,
4) residual phase noise after calibration is $30\degr$,
5) anomalous refraction.
The object is assumed to pass the meridian in the middle of a single observational run. We
use 30 minutes of integration time or increase it such that the deconvolved maps look similar to the input
model with chemical gradients. In contrast to modern interferometric facilities there will be no problem
to achieve a good $uv$ coverage with ALMA in a fraction of an hour.

\begin{figure}
\epsscale{1.02}
\plotone{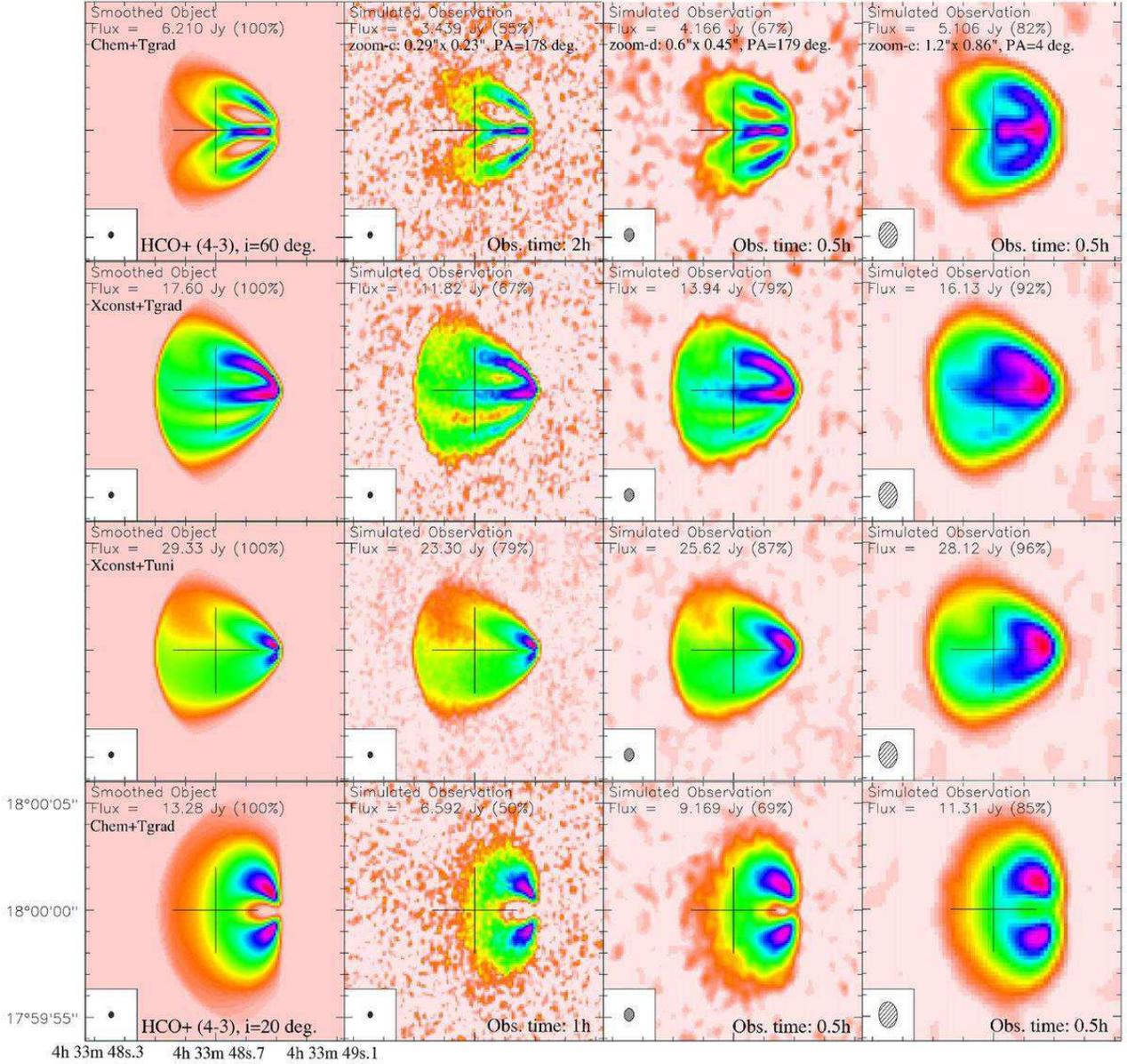}
\caption{(From first to third row) The same as in Fig.~\ref{chan_maps_3} but processed with the
GILDAS ALMA simulator for the three array configurations and 64-antennas:
``zoom-c'' ($\sim 0.25\arcsec$ beam size, integration time is 2~hours),
``zoom-b'' ($\sim 0.5\arcsec$ beam size, integration time is 0.5~hours),
and ``zoom-e'' ($\sim 1\arcsec$ beam size, integration time is 0.5~hours).
(Bottom row) The HCO$^+$(4-3) channel map at the $V=+0.3$~km\,s$^{-1}$ velocity channel
for the disk model with chemical gradients and the inclination angle of $20\degr$.}
\label{alma}
\end{figure}

The simulated ALMA channel maps of HCO$^+$(4-3) at various spatial resolutions for the three adopted models
and the inclination angle of $60\degr$ are presented in Fig.~\ref{alma} (upper 3 rows).
All significant features that are present in the ``ideal'' molecular emission spectra appear
clearly at the resolution of $0.25\arcsec$ (``zoom-c'' array configuration) and some are still visible at half
the resolution (``zoom-d'' configuration). The use of even lower resolution (``zoom-e'') makes it difficult
to disentangle the chemical and thermal features without extensive modeling.
The DM~Tau disk model with layered HCO$^+$
abundances and 2D temperature gradient has the lowest line intensities and hence requires the longest
observation of 2~hours, while for the models with uniform HCO$^+$ abundances it can be as short as 30 minutes or less.

In addition, we perform a similar analysis for a face-on orientation with $i=20\degr$.
At such low inclination channel maps for distinct disk models look similar and
multi-line, multi-molecule analysis become a must.
The corresponding $V=0.3$~km\,s$^{-1}$ channel of the HCO$^+$(4-3) channel map for the disk model
layered abundances is presented in Fig.~\ref{alma} (bottom row). This channel map has a
ring-like shape with two emission peaks located around the low-intensity midplane. The intensity in this channel is by
a factor of 2 stronger than for the disk model with $i=60\degr$ and thus the observational time required to detect
all these features is only 1~hour with the $0.25\arcsec$ beam size and less than 30 minutes at lower resolutions.
All major features in the $i=20\degr$ channel map are resolved even with the compact ``zoom-e'' array configuration.

Given the importance of multi-line observations and ability of ALMA to simultaneously observe several
transitions of several species in various frequency bands,
we continue our study for other observationally important molecular tracers (CS, HCN, CO isotopes)
as well as a smaller disk with a radius of $\approx 250$~AU. All results are summarized in Table~\ref{obs_conditions}.
The channel maps for other transitions at millimeter wavelengths and for molecules located in the
intermediate layer are similar to that of HCO$^+$, though their intensities are vastly different.
We find that smaller disks are better studied at high frequencies of 400--700~GHz with moderately extended
array configurations leading to smaller beam sizes than at millimeter wavelengths, albeit longer integration times
are required for that. The contamination of the high-lying emission lines by the optically thick dust continuum
from the disk inner regions and non-LTE excitation can be an issue for the analysis
of these high-frequency data.

The molecular lines also comprise kinematic information. High-quality
ALMA spectra will allow to derive accurate dynamical masses of the central stars and thus calibrate pre-main-sequence
evolutionary track models \citep{SDG00}, discern gas motions along spiral arms in gravitationally unstable disks
\citep{Fromangea04}, and trace flows of matter around a forming planet \citep{P07}. We do not pretend to analyze
the ALMA requirements necessary for such studies as it would deserve a separate publication.

%--------------------------------------------
\section{Conclusions}
%--------------------------------------------
We showed that the ALMA interferometer will allow
us to distinguish the effects of temperature gradients and chemical
stratification in disks through molecular line observations, in particular
for highly-inclined objects.
We found that moderately extended array configurations (with baselines of $\la 1$~km)
and 0.5--10 hours of integration time will be necessary to pursue such
observations.
%at millimeter wavelengths.

\acknowledgments
The authors are grateful to an anonymous
referee for valuable comments and suggestions.
This research has made use of NASA's Astrophysics Data System.
% {\it Facilities:} \facility{ALMA}
%%%%%%%%%%%%%%%%%%%%%%%%%%%%%%%%%%%%%%%%%%%%%%%%%%%%%%%%%%%%%%%%%%%
% References:
%%%%%%%%%%%%%%%%%%%%%%%%%%%%%%%%%%%%%%%%%%%%%%%%%%%%%%%%%%%%%%%%%%%

%\clearpage
%====== Fig. 1 ===========
%========================

%\clearpage
%====== Fig. 2 ===========
%[\textit{See the electronic edition of the Journal for an animation showing all channels.}]}
%========================

%\clearpage
%====== Fig. 3 ===========
%========================

%========================
% The requirements for ALMA to detect large-scale chemical structures in disks.
%\clearpage
\begin{deluxetable}{lccllll}
\tablewidth{0pt} \rotate %\tabletypesize{\footnotesize}
\tablecaption{Requirements for ALMA to study chemical and thermal gradients in disks\label{obs_conditions}}
\tablehead{
\colhead{Species} & \colhead{Frequency} & \colhead{Bandwidth} & \multicolumn{2}{c}{$R_{\rm disk}=800$~AU} &
\multicolumn{2}{c}{$R_{\rm disk}=250$~AU} \\
\colhead{} & \colhead{GHz} & \colhead{kHz} & \colhead{$i=20\degr$} & \colhead{$i=60\degr$} &
\colhead{$i=20\degr$} & \colhead{$i=60\degr$}
}
\startdata
HCO$^+$(1-0)         & 89 & 30   & zoom-c ($4^a$h)      & zoom-c ($10$h)     & zoom-a/b ($>12$h)  &  zoom-a/b ($>12$h)  \\
C$^{18}$O(2-1)       & 220 & 75  & zoom-d ($1$h)      & zoom-c ($<0.5$h)    & zoom-c ($4$h)      & zoom-c ($10$h)  \\
$^{13}$CO(2-1)       & 220 & 75  & zoom-d ($<0.5$h)   & zoom-d ($<0.5$h)    & zoom-c ($2$h)      & zoom-c ($3.5$h)  \\
CS(5-4)              & 245 & 80  & zoom-e ($3$h)      & zoom-d ($12$h)     &  zoom-b ($>12$h)    & zoom-b ($>12$h)  \\
HCN(3-2)             & 266 & 90 & zoom-e ($<0.5$h)    & zoom-d ($1$h)      & zoom-c ($4$h)      & zoom-b ($>12$h) \\
HCO$^+$(4-3)         & 357 & 120 & zoom-d ($<0.5$h)   & zoom-e ($<0.5$h)    & zoom-c ($2$h)      & zoom-c ($3$h) \\
HCO$^+$(7-6)         & 624 & 210 & zoom-e ($<0.5$h)   & zoom-e ($1.5$h)     & zoom-c ($12$h)     & zoom-d ($>12$h) \\
$^{13}$CO(6-5)       & 661 & 220 & zoom-e ($<0.5$h)   & zoom-e ($1$h)       & zoom-d ($1$h)     & zoom-c ($6$h)
\enddata
\tablenotemark{a}\tablenotetext{*}{Observing times are computed for the ALMA made of 64-antennas. With 50 antennas,
these values should be increased by a factor of 1.5.}
\end{deluxetable}

\end{document}